\documentclass[aps,prb,twocolumn,groupedaddress,showpacs]{revtex4-1}

\usepackage{bm}
\usepackage{graphics}
\usepackage{graphicx}
\usepackage{amsthm}
\usepackage{amsmath}
\usepackage{wasysym}
\usepackage{amssymb}
\usepackage[dvips]{epsfig}
\usepackage{color}

\begin{document}

\title{Exactly Solvable Topological Chiral Spin Liquid with Random Exchange}

\author{Victor Chua and Gregory A. Fiete}
\affiliation{Department of Physics, The University of Texas at Austin, TX 78721, USA}

\date{\today}

\begin{abstract}
We extend the Yao-Kivelson decorated honeycomb lattice Kitaev model [Phys. Rev. Lett. {\bf 99}, 247203 (2007)]  of an exactly solvable chiral spin liquid by including disordered exchange couplings.  We have determined the phase diagram of this system and found that disorder {\em enlarges} the region of the topological non-Abelian phase with finite Chern number.  We study the energy level statistics as a function of disorder and other parameters in the Hamiltonian, and show that the phase transition between the non-Abelian and Abelian phases of the model at large disorder can be associated with pair annihilation of extended states at zero energy.  Analogies to integer quantum Hall systems, topological Anderson insulators, and disordered topological Chern insulators are discussed.
\end{abstract}

\pacs{}
\pacs{75.10.Kt,73.43.-f,03.65.Vf}


\maketitle
\section{Introduction}
Quantum spin liquids in two dimensions are often difficult to firmly establish theoretically because of their strongly correlated nature \cite{Balents:nat10}.  An important advance in this direction was made by Kitaev who discovered a class of spin models with spin liquid ground states that can be solved exactly \cite{Kitaev:ap06}.  These models offer an existence proof for many proposed exotic spin liquid states, such as those with non-Abelian excitations \cite{Kitaev:ap06,Yao:prl07}.  A number of closely related models have been studied in two and three dimensions \cite{Yao:prl10}
, and controversial states such as such as two dimensional gapless spin liquids with a stable spin Fermi surface have been established \cite{Chua:prb11}.  Besides the existence of certain types of quantum spin liquids, their response to local impurities, and disorder more generally, is an important and largely unexplored issue \cite{Florens:prl06}.
To date, only a few studies of local defects and impurities in Kitaev models have been published \cite{Willans:prl10}.

In this article we examine an exactly solvable fully disordered version of a Kitaev model, Eq.\eqref{eqn:Ham}, on the decorated honeycomb lattice shown in Fig.\ref{Fig:lattice}. {As was originally suggested by Kitaev \cite{Kitaev:ap06} such a lattice with triangular plaquettes could lead to a chiral spin liquid ground state. In the clean limit, this model is known to possess both Abelian and topological (finite Chern number) non-Abelian gapped chiral spin liquid phases \cite{Yao:prl07,Dusuel:prb08,Kells:prb10}.}  In this work, we focus on how random exchange disorder affects the phase boundaries and show there are analogs to the recently studied topological Anderson insulators \cite{Jain:prl09,Groth:prl09,Prodan:prb11} and disordered Chern insulators \cite{Prodan:prl10}.  Our main result is that disorder {\em enlarges} the parameter space of the topological phase and therefore {\em can drive a transition into the topological phase}.  This implies that disorder could play a useful role in experimentally realizing certain classes of topological quantum spin liquids.  We study in detail the Abelian and non-Abelian phases, and the transition between them as a function of disorder by examining (i) the spatial distribution of thermal currents under a small temperature gradient and (ii) the energy level statistics.  We show that in the large disorder limit the transition between the two phases can be understood as the pair annihilation of extended bulk states at zero energy, analogous to the integer quantum Hall effect (IQHE). To the best of our knowledge, a study of random disorder in a chiral Kitaev model has not been carried out before.  

{The paper is organized as follows. In section II we introduce the Yao-Kivelson model, its mapping to non-interacting Majorana fermions and its different ground states. Next, we introduce the form of exchange disorder that we have studied. In section III we discuss how the Chern number of the disordered system is determined through the numerical calculation of the linear response thermal Hall conductivity. We then present the disordered phase diagram of this system. Next we visualized the local thermal currents in the topologically non-trivial phase at large disorder. In section IV we report the results of a energy level statistics analysis that is used to characterize the disordered Majorana mode wavefunctions at various points in the phase diagram. Finally, in section V we comment on the behavior of the spectral gap and thermal conductivity at large disorder and near the quantum phase transition.}

\section{Model}
\begin{figure}[t]
\includegraphics[width=7cm]{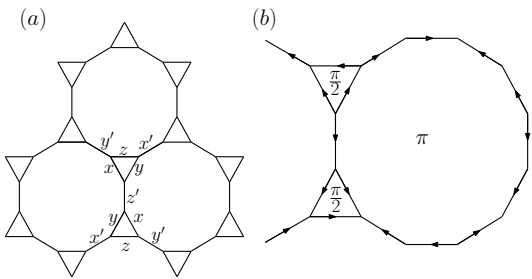}
\caption{(a) The decorated honeycomb lattice and the labeling of links describing the interactions in Eq.\eqref{eqn:Ham}. (b) The $\mathbb{Z}_2$ gauge field $u_{ij}$ configuration and their ground state fluxes \cite{Yao:prl07}.\label{Fig:lattice}}
\end{figure}

We consider spin-1/2 moments on the lattice shown in Fig.\ref{Fig:lattice} with Hamiltonian,
\begin{eqnarray}
H= \sum_{x-\mbox{link}}J\sigma_i^x\sigma_j^x
+\sum_{y-\mbox{link}}J\sigma_i^y\sigma_j^y
+\sum_{z-\mbox{link}}J\sigma_i^z\sigma_j^z \nonumber \\
+\sum_{x'-\mbox{link}}J'\sigma_i^x\sigma_j^x
+\sum_{y'-\mbox{link}}J'\sigma_i^y\sigma_j^y
+\sum_{z'-\mbox{link}}J'\sigma_i^z\sigma_j^z ,\;\;\;\;
\label{eqn:Ham}
\end{eqnarray}
where $\mu=x,y,z,x',y',z'$ denotes the various types of links in the lattice, labelled as in Fig.\ref{Fig:lattice}  The $J,J'>0$ are exchange couplings between the spins represented by the Pauli matricies $\sigma_j^a$, $a=x,y,z$ on site $j$.  The presence of plaquettes with an odd number of sides results in the spontaneous breaking of time-reversal symmetry \cite{Kitaev:ap06,Yao:prl07}. 

The Hamiltonian \eqref{eqn:Ham}  is solved by introducing Majorana operators $\xi_i^x,\xi_i^y,\xi_i^z,c_i$ at each site $i$ \cite{Kitaev:ap06}. The combination $\sigma_i^\mu\equiv i\xi_i^\mu c_i$ faithfully reproduces the Pauli operator algebra provided $\xi_i^x\xi_i^y\xi_i^z c_i=1$ is enforced with  projector $P=\prod_{i}\left( \frac{1+D_i}{2}\right)$ where $D_i=\xi_i^x\xi_i^y\xi_i^z c_i$. Recasting (\ref{eqn:Ham}) in terms of Majoranas, yields a ``tight-binding" Hamiltonian for $c$ in a static background $\mathbb{Z}_2$ gauge field 
\begin{equation}
\widetilde{H}=\sum_{\left< ij \right>\in \mu-\mbox{links}}  J \,i \,u_{ij} c_i c_j
+\sum_{\left< ij \right>\in \mu'-\mbox{links}}  J' \,i \,u_{ij} c_i c_j,
\label{eqn:HamMaj}
\end{equation}
where $u_{ij}=-i\xi_i^\mu \xi_j^\mu$ with $\mu=x,y,z$ and $\mu'=x',y',z'$. $P\widetilde{H}P=H$ then recovers (\ref{eqn:Ham}). The $u_{ij}$ commute with $\widetilde{H}$ and thus provide good quantum numbers $u_{ij}=-u_{ji}=\pm 1$. They are non-dynamical $\mathbb{Z}_2$ gauge fields. Thus, their fluxes are gauge invariant and integral invariants of motion of $H$.  The $\widetilde{H}$ ground state (which is degenerate) is specified by the flux configuration which has the least energy when no excitations are present. Yao and Kivelson \cite{Yao:prl07} have shown that the minimal flux configuration is the one illustrated in Fig.\ref{Fig:lattice} (b). In the clean system, the ground state of $\widetilde{H}$ for $J'/J<\sqrt{3}$  is gapped and topologically non-trivial with Chern number $\nu=\pm 1$, while for $J'/J>\sqrt{3}$ the ground state is gapped with $\nu=0$ \cite{Yao:prl07}. 

{It was discussed in ref. \cite{Yao:prl07} that within a flux sector, $P$ projects out half the possible eigenstates of $\widetilde{H}$ and this is believed to be a general feature of all Kitaev models \cite{Yao:prl07,Yao:prl09, Pedrocchi:prb11}.} In particular, the physical states will depend on the relative sign of the exchange couplings $J$ and $J'$, the global boundary conditions ({\it e.g.} on a torus), the specific flux sector and whether the number of $c$-field excitations is even or odd.  We have verified our ground states are physical by numerically applying the projection protocol of Ref.\cite{Pedrocchi:prb11} to the Yao-Kivelson model for finite system sizes. We then considered a disordered version of \eqref{eqn:Ham} by allowing $J$ and $J'$ to vary randomly from link to link as
\begin{eqnarray}
&&J_{<\text{ij}>}=j\left(1+X_{<\text{ij}>}\right), \;\;\; J'_{<\text{ij}>'}=j'\left(1+X_{<\text{ij}>'}\right), 
 \label{eqn:disH}
\end{eqnarray}
where $<ij>$ are any of the $x,y,z$-links and $<ij>'$ any of the $x',y',z'$-links. The random variables  $X_{<\text{ij}>}$ have uniform distribution $P\left( \left|X_{<\text{ij}>}\right| \leq W/2 \right)=1/W$ and zero otherwise. The parameters $j$ and $j'$ set the average; $W$ is dimensionless and is limited to $W<2$ so that $J,J' >0.$  By adiabatic continuity, this guarantees \eqref{eqn:HamMaj} remains in the ground state flux sector shown in Fig.\ref{Fig:lattice} (b).

\section{Thermal Conductivity}
We consider a finite rectangular system with $18\times 27$ unit cells (each with with 6 spins) with $N=2196$ sites and open boundary conditions. For a finite system with $\nu=\pm 1$ for some choice of $j,j'$ and disorder realization, an edge mode should be present at zero energy in the bulk gap \cite{Yao:prl07}.  In contrast to the IQHE, these states do not couple to conventional electromagnetic fields, but nevertheless do coupled to thermal gradients. The topological phase with  $\nu=\pm 1$  would be manifested as a quantized thermal Hall conductivity, $\sigma_{xy}^\text{th}$, with $\sigma_{xy}^\text{th} =\frac{\pi}{24 \hbar} k_B^2 T \nu$, where $k_B$ is Boltzmann's constant and $T$ the absolute temperature \cite{Kitaev:ap06}.  We computed
$\sigma_{xy}^\text{th}$  numerically for specific disorder realizations and then averaged over 100 realizations. 

\begin{figure}[th]
\includegraphics[width=\columnwidth]{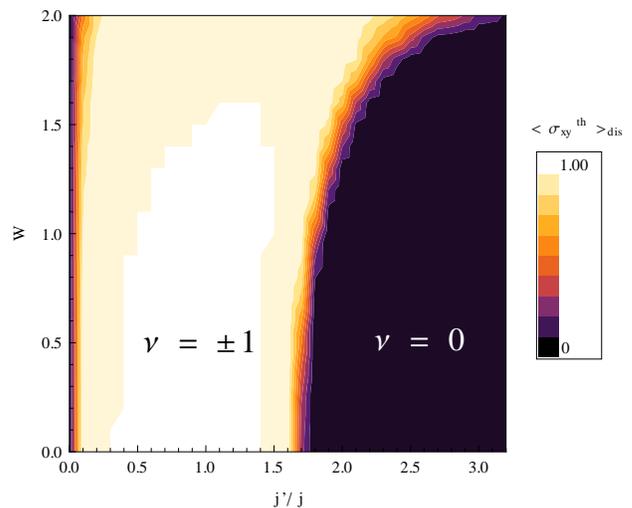}
\caption{(Color online) A contour plot of $\langle \sigma_{xy}^\text{th}\rangle_\text{dis}$ in units of $\frac{\pi}{24\hbar} k_B^2 T$. Finite size effects lead to a rounding and a small shift in the critical $j'_{\text{crit}}=\sqrt{3}j$ expected in the clean limit for the transition between topological and non-topological phases \cite{Yao:prl07}.\label{Fig:conductivity}}
\end{figure}
\begin{figure*}[th]
\includegraphics[width=2\columnwidth]{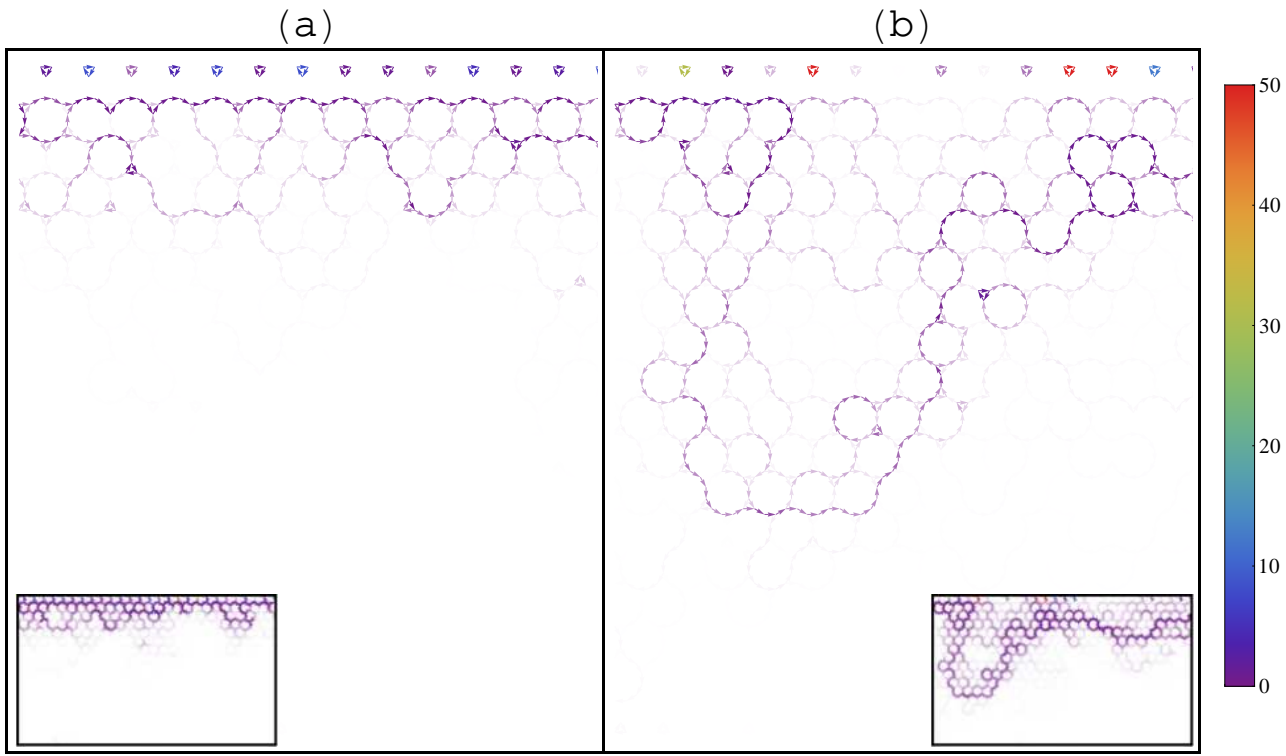}
\caption{(Color online) Local thermal current density profiles, $I^{\text{th}}_{ij}$, of a disordered quasi one-dimensional ``wire" (running right to left) for typical realizations possessing chiral edge modes at large disorder $W=1.75$. (a) $j'=1.5j$ (b) $j'=2.1j$.   The currents are displayed in units of  $\frac{\pi}{24\hbar}k_B^2 T$. Currents below 0.5 are shown as translucent and vary from fully transparent $(I^\text{th}_{ij}=0)$ to fully opaque $(I^\text{th}_{ij}=0.5)$. Only a part of the system is shown in fine detail, but the respective insets contain of coarse grained current profiles of the entire ``wire" lattice between the two ``leads".}\label{Fig:Currents}
\end{figure*}

Since (\ref{eqn:HamMaj}) with \eqref{eqn:disH} is exactly diagonalizable, the conductivity can be determined within the Keldysh Green's function formalism which has been successfully applied to mesoscopic transport in quantum wires \cite{Cresti:2003,Metalidis:2005,Jain:prl09}. We implemented this technique by coupling our disordered system to two ideal ``leads" which are modeled as clean versions of the system with $j'=j$ such that chiral edge modes are supported in the leads. {If the disordered system under study possesses chiral edge modes, these and only these states will couple perfectly (without reflection) with the modes already present in the leads at low energies. Thus a calculation to linear response of the thermal conductance, will then yield the quantized conductance. If however the system does not possess chiral edge modes and has $\nu=0$ which is different form that of the ideal leads, the thermal transmission will be heavily or entirely suppressed. We should also point out that if the leads do not support low energy chiral edge modes, then the thermal conductance will always be trivially zero. One could also use a different system altogether to model the leads, say the honeycomb Kitaev model in its gapped chiral phase. The crucial requirement is that it is gapped in the bulk but has edge modes of the same number and chirality as the modes of the clean system in the non-Abelian phase.} 

The right lead is then set to $T=0$ and the left is kept at a very small $T>0$. To linear order in $T$, $\sigma_{xy}^\text{th}$ is given by a Chester-Thellung-Kubo-Greenwood formula \cite{Chester:1961,Cresti:2003,Metalidis:2005,Jain:prl09},
\begin{equation}
\sigma_{xy}^\text{th}=\frac{\pi  k_B{}^2T}{24\hbar}\text{tr}\left\{\Gamma ^{\text{left}}G^{\text{ret}}\Gamma ^{\text{right}}G^{\text{adv}}\right\}|_{E =0},
\end{equation}   
where the single-particle retarded and advanced Green's functions are evaluated at zero energy ($E=0$) and the trace is taken over transverse modes. The leads, which serve as a heat source and sink, are modeled by self energy matrices, $\Sigma^{(\text{left}/\text{right}),(\text{ret}/\text{adv})}$, which only act on sites that interface with the leads. These extra terms are included in the calculation of $G^{\text{ret}/\text{adv}}$, and $\Gamma^{\text{left/right} }$ are associated with the imaginary parts of $\Sigma^{(\text{left}/\text{right}), \text{ret}}$.  The results of these calculations yielded the conductivity contour plot in Fig.\ref{Fig:conductivity}, which also serves as a phase diagram.  The most striking feature is the expansion of the topologically non-trivial phase region for intermediate to strong values of disorder strength, $1<W<2$.  Similar effects have also been observed in disordered topological insulators \cite{Jain:prl09,Groth:prl09,Prodan:prb11} and disordered Chern insulators \cite{Prodan:prl10}.

We visualized the local thermal current density across the link $<ij>$, $I^{\text{th}}_{ij}$, in Fig. \ref{Fig:Currents} for large $W$ disorder realizations in the non-trivial topological phase $\nu=\pm 1$ by adapting the methods of Ref.\cite{Metalidis:2005}. The $I^{\text{th}}_{ij}$ are determined from the non-equilibrium Green's function $G^{<}$ in the linear response regime of small $T$.  Fig.\ref{Fig:Currents}(a) clearly demonstrates the chiral nature of the edge modes which are localized on the upper edge. However, as $j'$ is tuned towards criticality as in Fig.\ref{Fig:Currents}(b), the edge currents begin to impinge into the bulk. Eventually a backscattering channel opens when the currents connect the different edges. In close analogy to the plateau transitions of the IQHE, the topological phase transition can be associated with the closure of the bulk gap and the appearance of an extended bulk eigenstate at $E=0$ that percolates the system. 

\begin{figure*}
\includegraphics[width=\textwidth]{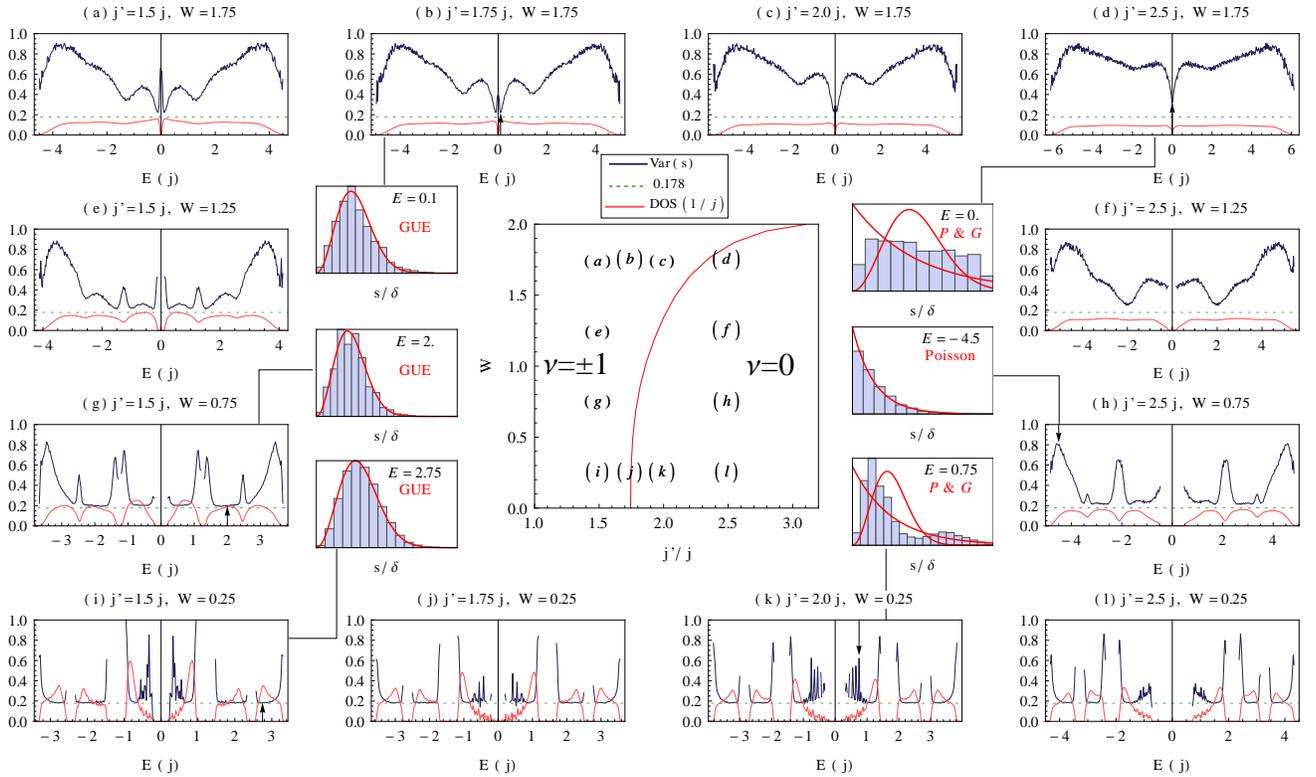}
\caption{(Color online) Energy level statistics at various points in the phase diagram. (a)-(l) The variance of the energy spacing $s$ shown as the solid blue (dark gray) line in units of mean level spacing $\delta$ against the GUE expectation of 0.178 shown as the horizontal dashed green line. Also shown is the averaged density of states which is the solid red (light gray) line. Statistics are derived from $10^3$ disorder realizations and binned in energy windows of width $\Delta E=0.25\,j$. Shown in the center are the locations of (a)-(l) in an approximate phase diagram. Spacing distribution histograms are taken from energy windows (indicated by the black lines and arrows) of special interest with comparisons to the Wigner GUE and Poisson distributions. The finite gap at $E=0$ does not close in (a)-(c) which is not apparent here with this scale and bin size.}
\label{Fig:ELS}
\end{figure*} 

\section{Energy Level Statistics}
{The results of the previous section suggest that even at large finite disorder, the phase transition in the \emph{bulk} proceeds from a thermal insulator to a thermal insulator via a bulk thermal metal at criticality. This critical metallic phase is manifested by the emergence of an extended Majorana mode in the bulk at $E=0$. Thus with the aim of determining the localized/extended character of the Majorana mode wavefunctions, we have performed an energy level statistics analysis with comparisons to the expectations of random matrix theory. More specifically, a statistical analysis of closest energy level spacings, $s$, allows for an indirect determination of the spatial character of the wavefunctions across the spectrum \cite{Prodan:prl10, Shklovskii:1993}. In addition an analysis based on level spectra has the technical advantage of not requiring the explicit computation of disordered wavefunctions. Hence it has allowed us to efficiently study the entire spectrum for many points on the phase diagram without the need for intensive computing resources.} Since $\widetilde{H}$ possesses particle-hole but no other symmetries, it falls into the $D$-symmetry class of the ten-fold Altland and Zirnbauer classification \cite{Altland:1997}, which is also shared by Bogoliubov-de Gennes (BdG) Hamiltonians with the same symmetries. Moreover, it is predicted that at energies $E$ much greater than the mean level spacing, $\delta$, the energy level statistics will reduce to that of the Gaussian Unitary ensemble (GUE) of Wigner and Dyson \cite{Altland:1997}. Delocalized eigenstates of similar energies are then expected to experience level repulsion described by the GUE Wigner surmise, $P(s)=\frac{32}{\pi^2}s^2 e^{-4s^2/\pi}$, where $s$ is measured in units of $\delta$. Level repulsion is exemplified by the decay of probability density as $s\rightarrow 0^+$. In contrast, strongly localized eigenstates of similar energies experience less level repulsion and are more likely to be described by Poisson statistics $P(s)=e^{-s}$. By gathering statistics of level spacings from energy windows across the spectrum and making comparisons to these limits, one can distinguish between regions of strong localization, intermediate localization, and delocalization. We have performed such an analysis, and a summary of our results is presented in Fig.\ref{Fig:ELS}.  Spectra from disordered samples for systems of the dimensions mentioned earlier with periodic boundary conditions at various values of $j'$ and disorder strength $W$ were averaged over $10^{3}$ realizations.  Shown in Fig.\ref{Fig:ELS}(a-l) are the variances of the normalized energy level spacings $\text{Var}(s)$ and the averaged density of states (DOS). 

For small disorder $W=0.25$ the bands are fairly well defined by clear energy gaps. Within these bands lie regions of extended states distinguished by plateaus where the spacing variances agrees well with the GUE value 0.178. Samples taken from these low regions of variance also indicate a strong agreement with the GUE distribution. Towards the band edges however, the distribution of spacings tend to resemble Poissonian statistics, and more crucially the level repulsion is greatly reduced. With increasing $W$, the energy bands broaden and close some gaps in the DOS but there still remains generically a strong repulsion from $E=0$ except at a phase transition. As $W$ increases, the spacing variances across the spectrum increase and exhibit less level repulsion. Interestingly, the DOS and spacing variances on either side of the critical line behave rather differently. Specifically, the topological $\nu=\pm 1$ phase develops a narrow robust region of extended-like states with strong level repulsion located near $E=0$ which is marked by a clear dip in the spacings. As shown by the histogram taken from a bin in Fig.\ref{Fig:ELS}(b), states in this energy window exhibit strong level repulsion which matches the GUE distribution fairly well. By increasing $j'$ in Figs.\ref{Fig:ELS}(a-d), we also observed that these extended states eventually merge or annihilate at the phase transition. These results then allow us to connect with the findings of the previous section. Namely, at the phase transition there emerges an extended $E=0$ state(s) which we associate with the pair annihilation at $E=0$ of two extended regions which evolve from the bulk bands. We note that an identical phenomena of pair annihilation had also been previously observed in topological insulator, IQHE and topological superconductor systems \cite{Prodan:prl10,Onoda:2007}. This system, however, is the first example of a disordered Kitaev model shown to exhibit this sort of physics.  We remark that Fig.\ref{Fig:ELS}(d) indicates an intermediate distribution near the transitions, and Fig.\ref{Fig:ELS}(k) shows an interesting bimodal distribution in the vicinity of strong DOS fluctuations.

\begin{figure}
\includegraphics[width=\columnwidth]{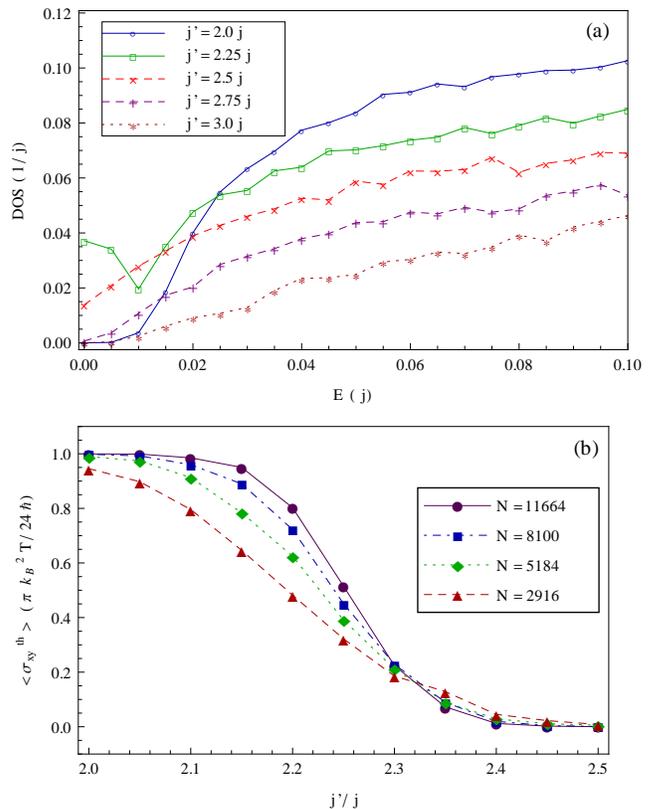}
\caption{{(Color online) Spectral and transport properties in the vicinity of the phase transition at large disorder strength $W=1.75$. (a) Enlarged plots of the disorder averaged density of states (DOS) near $E=0$ where $j'$ varies between the topologically non-trivial phase ($2.0j$) to the topologically trivial ($3.0j$) for a system of fixed size $N=2916$. Each point was derived from $10^3$ realizations. (b) The disorder averaged thermal Hall conductances $<\sigma_{xy}^{th}>$ across the phase transition and for various system sizes $N$ but constant aspect ratio. Conductances are expressed in the quantized thermal conductance unit $\frac{\pi}{24\hbar}k_B^2 T$ and were averaged over 500 realizations. The critical $j'$ is inferred to be $j'_{\text{crit}}\approx 2.31j$}}
\label{Fig:Gap}
\end{figure} 

\begin{figure}
\includegraphics[width=\columnwidth]{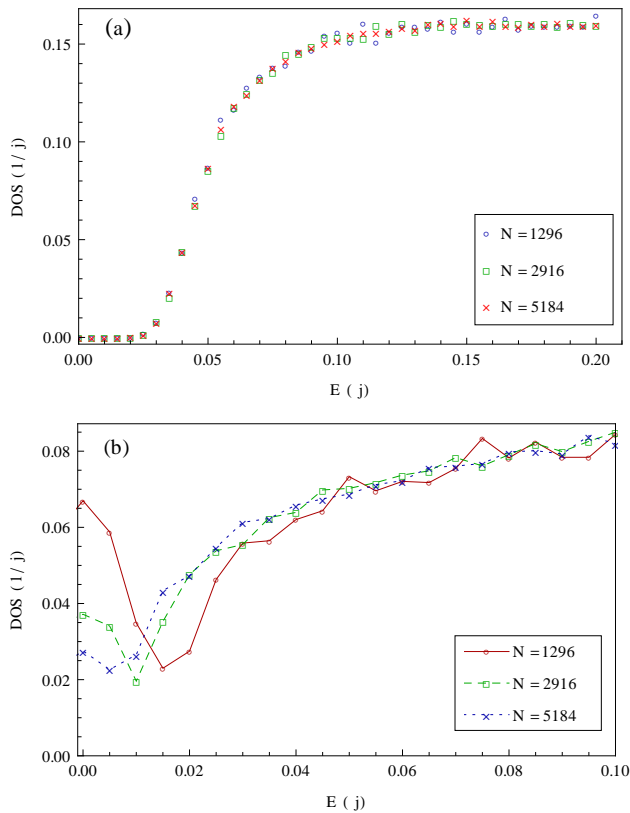}
\caption{{(Color online) Disorder averaged density of states (DOS) near $E=0$ for various system sizes and large disorder strength $W=1.75$. (a) Deep in the topological $\nu=\pm1$ phase $j'=1.5j$ where a robust gap develops which is size insensitive. (b) In proximity of the phase transition $j'=2.25j$ where the gap closes but $\text{DOS}(E=0)$ diminishes with increasing system size.}}
\label{Fig:Gap2}
\end{figure} 

\section{Large Disorder and Finite Size Effects}
{In this last section we briefly comment upon the spectral and transport properties of the most interesting part of the phase diagram. That is at large disorder and across the phase boundary. We also discuss related finite size effects that are observed. 

First we study the gradual closing of the gap at large constant disorder strength. An enlarged plot of the disorder averaged density of states (DOS) near $E=0$ around the transition is shown in Fig.\ref{Fig:Gap}(a). Spectra was taken from a system of size $N=2916$ and disorder strength $W=1.75$ with $j'$ varying between $2.0j$ and $3.0j$. For these parameters and to within our numerical precision, the gap is seen to close significantly in a window between $j'=2.25j$ and $j'=2.50j$. For comparison, the disorder averaged thermal Hall conductivities were computed for the relevant range of parameters and plotted in Fig 5(b). The conductivities of larger system sizes was also studied and is also shown in Fig.\ref{Fig:Gap}(b). The additional data allows us to put the critical $j'$ at approximately $2.31j$. This is the point where all the conductivity curves cross.   
 
Next we examined the DOS near $E=0$ for various system sizes near and away from the transition. Shown in Fig.\ref{Fig:Gap2} are DOS plots near $E=0$ for system sizes $N=1296,2916$ and $5184$. We remind the reader that the system size used in our main results is $N=2916$ sites and we have not spectrally studied larger sizes in as much detail due to current hardware limitations. Interestingly the system size dependence is affected by how close $j'$ is to its critical value. Deep in the topological $\nu=\pm1$ phase with $j'=1.5j$ as shown in Fig.\ref{Fig:Gap2}(a), the DOS varies weakly with system size near the gap. However in Fig.\ref{Fig:Gap2}(b), when $j'=2.25j$ and where the gap is seen to close, the average density of states at $E=0$ diminishes noticeably with increasing system size. It suggests that the gap's closure at $j'=2.25j$ is really a finite size effect. Although, we believe that a more detailed spectral analysis with larger samples and sizes is required to firmly establish this, which is planned for future work.    

Thus we have confirmed that for a finite system the gap is closed at the transition. However, to within our numerical precision it remains closed over a finite window of parameters. But this window lies inside a region of rapidly changing conductivity which contains the critical $j'$. Furthermore, we believe this behavior of the gap to be a finite size effect. 

Finally we remark that while it is certainly necessary that the gap closes at the transition, it may not be a sufficient condition. In a disordered system where disorder broadening may occur, the gap may be closed by states that are localized but are not topological. In an IQHE system, this would correspond to the Fermi level being placed in a band of localized states. Nevertheless, our numerics indicate that at least for our choice of model disorder and at large disorder strengths, the E=0 gap remains robust in the topological phase and gradually closes at the transition, but not sharply.}
\newline

\section{Conclusion}
We report a disordered generalization of the exactly solvable Yao-Kivelson-Kitaev model which supports chiral spin-liquid ground states with both trivial $\nu=0$ and non-trivial $\nu=\pm 1$ topological phases. We mapped out a disorder phase diagram of this system and showed that exchange disorder can enlarge the region of the topological phase. We visualized the local thermal current density of the topological phase in contact with heat reservoirs. A statistical analysis of the energy level spacings at various regions of the phase diagram yielded evidence for the nature of the wavefunctions in agreement with the thermal current studies. {The low energy gap was shown to be robust in the non-Abelian phase even at large disorder and closes at the onset of the phase transition.} Our work shows there are many interesting parallels between topological quantum spin liquids and other topological phases \cite{Fiete:pe11}, and we expect more to be discovered in the future.
    

{\it Acknowledgements}--This work was supported by ARO grant W911NF-09-1-0527 and NSF grant DMR- 0955778. V.C. thanks A. R\"uegg, M. Kargarian and V. Zyuzin for early discussions. 
\providecommand{\noopsort}[1]{}\providecommand{\singleletter}[1]{#1}%

\end{document}